\documentclass[preprint,aps,amsmath]{revtex4}
\usepackage{graphicx}

\begin{document} 

\title{ZM theory V: Lorentz force equation and the vector potential}

\author{Yaneer Bar-Yam}

\affiliation{New England Complex Systems Institute, Cambridge, Massachusetts 02138}

\begin{abstract}

In ZM theory the direction of time has a non-zero projection onto space and this projection corresponds to the local velocity relative to the observer. Classical trajectories can be obtained by following the local direction of time. The relationship of time to space enables the change in momentum over time to be related to the spatial change in energy and momentum. Previously Hamilton's equations-of-motion were derived by considering trajectories in one space and one time dimensions. Here we consider three space and one time dimension. Without any other assumptions we derive the Lorentz force law of electromagnetism with relevant definitions of the scalar and vector potentials. 

\end{abstract}


\maketitle

ZM theory posits a cyclical ``clock" field whose space-time variation determines the local definition of time as seen by an observer. The direction of local time is not orthogonal to space, and the projection of the direction of time onto space is the local coordinate system velocity.\cite{ZM1,ZM2,ZM3,ZM4,ZMQ} Defining the direction of time in terms of the variation of the local field direction leads to a Lorentz invariant theory which has a first order ``bent" space-time metric.\cite{ZM1} 
The clock field is associated with the description of a quantum particle described by the Dirac equation\cite{ZM4} or to a classical particle described by special relativity.\cite{ZM1,ZM2} The clock rate $m$ is the rest mass of the particle. 

It has also been shown that considering a 1+1 dimensional space-time, it is possible to follow the direction of time to define a trajectory, yielding the relativistic versions of Lagrange's and Hamilton's classical equations of motion.\cite{ZM2} In deriving Hamilton's equations of motion, it is necessary to define a generalized momentum and velocity dependent Hamiltonian $H(p,v)$. In ZM theory this is none other than the rate of rotation of the clock field along an arbitrary direction of time as specified by the velocity $v$. Determining the actual velocity by maximizing the variation of the clock field in the direction of time provides a variational principle which leads to the momentum dependent Hamlitonian $H(p)$, and Hamilton's equations of motion. In particular, the force equation results from relating $dp/dt$ to $dp/dx$, i.e. the time dependence to the space dependence, which results from the projection of the time direction onto space. In deriving Lagrange's and Hamilton's equations of motion, the potential energy is inferred. The negative of the spatial variation of the potential energy is equal to the change of the momentum with time as in the classical force equation. Remarkably, this treatment seems to provide a fundamental origin for the bizarre mathematics of Hamilton's principle\cite{Goldstein} in which velocity and momentum are treated as independent even though they are fixed in their relationship according to conventional mechanics.

Here we generalize these derivations to 3+1 dimensional space-time and find that the resulting force equations are the Lorentz force equations with appropriately defined scalar and vector potentials. Our derivation suggests that these are the only possible trajectories in ZM theory based upon a single cyclical clock field. 

\section{Clock field}

We review the basic framing of ZM theory from the first paper in the series ZM1\cite{ZM1}. We consider a system, $Z$, with some set of distinctly labeled states that perform sequential transitions in a cyclic pattern, i.e. an abstract clock (similar to a conventional ``non-digital" clock consisting of a numbered dial, here with a single moving hand). Discreteness of the clock will not enter into the discussion in this paper. The clock states can therefore be extended to cyclical continuum, $U(1)$. The state change of the clock defines proper time, $\tau$, as defined by the clock. The clock phase is
\begin{equation}
c(\tau) = m \tau  \mod 2\pi
\end{equation}
where 
\begin{equation}
m = 2\pi /T
\end{equation}
is the cycle rate in radians, and $T$ is the cycle period. Since the clock is cyclical it is also possible to represent the changing state using an oscillator language: 
\begin{equation}
\psi  = \exp ( - im \tau ).
\label{psi} 
\end{equation}
Since the clock phase is not analytic, derivatives should be defined in terms of $\psi$. However, locally with proper choice of the location of the discontinuity, or where analytic continuation is valid, derivatives can be defined in terms of $c$.

The units we use ultimately will correspond to taking the speed of light, and reduced Planck's constant, $\hbar$, to be one. This implies that mass, energy and frequency are measured in the same units. The notation is chosen anticipating that $m$ will become the `rest mass' of the clock when it is reinterpreted as a particle. 

To introduce the space manifold, $M$, in this paper we consider a three dimensional space manifold, where real valued parameters 
\begin{equation}
\begin{array}{ll}
\vec{x}&=(x_1,x_2,x_3) \\
&=x_1 \hat x_1+x_2 \hat x_2+x_3 \hat x_3
\end{array} 
\end{equation}
are associated with the environment. We will commonly write the set of parameters indexed by $i$, i.e. $x_i$ and omit the subscript in arguments of functions. We avoid vector notation until the conventional use of vector notation in electromagnetic theory makes it helpful for comparison with traditional expressions. For convenience, each of the $x_i$ parameters is measured in the same units as $\tau$. 

Properties of the environment may lead to variation of the observed clock state with $x_i$. The analytic and non-analytic properties of the clock field are important for understanding ZM theory. In this paper, we assume only that in a neighborhood of a point on the manifold incremental changes of the clock field are well defined. The variation of the clock field along the clock dimension is fixed at the particle mass, $m$, and the variation along the space dimensions is defined to be the momentum, $p_i$:
\begin{equation}
\begin{array}{ll}
m &= \partial_\tau c  \\
p_i &= -\partial_i c 
\end{array}
\end{equation}
The next section defines the direction of time and the variation of the clock along the direction of time, the Hamiltonian.

\section{Hamiltonian in three plus one dimensions}

We briefly review the derivation of the direction of time and the Hamiltonian in three plus one dimensions.\cite{ZM4} The derivation follows from choosing the direction of time as the space-clock direction of maximal rate of change of the clock field. The non-orthogonality of space and time allows us to write the time dependence of the clock (the energy) in terms of its spatial dependence (the momentum). This dependence is the Hamiltonian. 

Let $\hat x_i$ be the spatial coordinate directions, and $\hat \tau$ be the internal clock dimension, which is orthogonal to space. Consider possible directions of time $\hat s$ (unit vectors) in the Euclidean space consisting of space and $\tau$ dimensions:
\begin{equation}
\hat s =  v_\tau \hat \tau - \sum_i v_i \hat x_i , 
\end{equation}
where $v_i$ and $v_\tau$ are respectively the negative and positive directional cosines of the time direction along the $\hat x_i$ axis and $\hat \tau$ axes respectively. The choice of negative signs for the space direction corresponds to the conventional choice of a positive velocity of the system with respect to the observer for positive values of these variables. The directional cosines satisfy the normalization constraint:
\begin{equation}
v_\tau^2 +  \sum_i v_i^2=1.
\end{equation}

The variation of the clock along the time dimension is defined to be the energy, $H$:
\begin{equation}
H = d_s c
\end{equation}
The generalized Hamiltonian for arbitrary direction of time can be obtained by adding the variation along each of the orthogonal coordinate axes multiplied by the directional cosines: 
\begin{equation}
H = v_\tau \partial _{\tau}  c - \sum_i  v_i \partial _{i} c  
\end{equation}
Substituting the definition of $p_i$, $v_i$ and $m$ gives
\begin{equation}
H(p,v) = \sum_i p_i v_i + m v_\tau.
\label{Hequation}
\end{equation}
Substituting for $v_\tau$ using the normalization constraint on the velocity yields:
\begin{equation}
H(p,v) = \sum_i p_i v_i + m\sqrt {1 - \sum_j v_j^2 }.
\end{equation}
This corresponds to the relativistic Hamiltonian as a function of momentum and velocity in variational mechanics.\cite{Goldstein}

For a single clock, we determine $v_i$ and $v_\tau$ by maximizing the rate of change of the clock in time (in reference \cite{ZM4} the same result is obtained using a Lagrange multiplier rather than inserting the velocity constraint). Setting
\begin{equation}
\partial H(p,v) /\partial v_i = 0
\label{maximize}
\end{equation}
gives
\begin{equation}
\begin{array}{ll}
v_i &= - \partial_i c / \omega \\
&= p_i/\omega
\end{array}
\end{equation}
where 
\begin{equation}
\omega = \sqrt{m^2+\sum_i (\partial_i c)^2}.
\end{equation}
Substituting $v$ into $H(p,v)$ gives:
\begin{equation}
H(p) =  \omega(p) = \sqrt{m^2+\sum_i p_i^2}.
\end{equation}
which is the Hamiltonian as a function only of momentum. The correspondence of $v_i$ to the velocity results from defining the meaning of a trajectory in the following section. 

\section{trajectories}

In this paper we consider the possibility of a spatial variation in momentum. In classical physics we consider momentum to depend on time rather than on space. Here this time dependence arises from the spatial dependence of the momentum (similar to the quantum concept of momentum as a spatial operator) and the non-orthogonality of space and time. 

The direction of time specifies the movement of the observer's coordinate system relative to the locally defined fixed position, rather than the movement of a location of space. Thus, if the direction of time specifies the movement of the coordinate system to the left, this is seen as the movement of the position of an entity (part of the cyclical field) to the right. Hence, the appearance of a negative sign in defining the velocity relative to the space component of the time direction. This can be shown formally either by considering the displacement of the coordinate system origin, or considering the transformation of the spatial argument of the field function. Taking the first approach, at every location $x_i$, at time $t$, the observer coordinate system origin follows the local direction of time to yield a displacement of the field feature location $\widetilde{x}_i(t)$ according to the observer described by 
\begin{equation}
\frac{d \widetilde{x}_i(t)}{dt} = \frac{d}{dt} (x_i-\bar{x}_i(t))
\end{equation}
where $\bar{x}_i(t)$ is the origin of the coordinate system of the observer. By geometry of the angle of time relative to space, this gives 
\begin{equation}
d\widetilde{x}_i(t)/dt = - d \bar{x}_i(t)/dt = v_i(x) = p_i(x)/\omega
\end{equation}

We can write the time dependence of the clock field coordinate as well as space coordinate along the trajectory as:
\begin{equation}
\begin{array}{ll}
v_\tau &= d_t \tau  \\
v_i &= -d_t x_i  
\end{array}
\end{equation}
This enables us to write the Hamiltonian using the chain rule for field variation along the trajectory, 
\begin{equation}
H = \sum_i \partial _{i} c d_t x_{i} + \partial _{\tau}  c d_t \tau ,
\end{equation}
which, upon substitution, gives the same expression as Eq. (\ref{Hequation}).

\section{Hamilton's first equation in 3+1 dimensions}

The momentum, $p_i$, and velocity, $v_i$, both describe the relationship between the field and the observer. $p_i$ plays the role of an extrinsic parameter and $v_i$ is determined by it through a variational principle, and thus can be considered a function $v_i(p_j)$. Considering $p_i$ to be variable we can therefore write  
\begin{equation}
\begin{array}{ll}
 dH(p)/dp_i &= \partial H(p,v)/\partial p_i + \sum_j (\partial H(p,v)/\partial v_j)(dv_j/dp_i) \\
 &= \partial H(p,v)/\partial p_i \\
 &= v_i,
\end{array}
\end{equation}
where the first equality is the chain rule, the second equality arises from the variational time direction determination as given by 
Eq.(\ref{maximize}), and the final equality from the explicit form of $H(p,v)$. This is the first of Hamilton's equations. 

Since we have the explicit form of $H(p)$, we can also obtain the final expression by taking the derivative directly: 
\begin{equation}
\begin{array}{ll}
 dH(p)/dp_i &= p_i / \sqrt{ \sum_j p_j^2 + m^2}  \\
 &= p_i/\omega  \\
 &= v_i
\end{array}
\end{equation}

\section{Hamilton's second equation in 3+1 dimensions}

Hamilton's second equation of motion relates the time derivative of the momentum along a trajectory to the space derivatives of various quantities. In order to derive it we compare
\begin{equation}
\begin{array}{ll}
dp_i/dt &= \sum_j (\partial_j p_i)(dx_j/dt) \\
&=  \sum_j v_j (\partial_j p_i)  
\label{first}
\end{array}
\end{equation}
with
\begin{equation}
\begin{array}{ll}
d_i H(p) & = \sum_j (dH/dp_j) \partial_i p_j \\
& = \sum_j v_j  (\partial_i p_j ) 
\label{second}
\end{array}
\end{equation}
In one dimension these are the same and the derivation is complete \cite{ZM2}. Here we subtract them to obtain:
\begin{equation}
\begin{array}{ll}
dp_i/dt - \partial_i H(p) &=  \sum_j v_j (\partial_j p_i - \partial_i p_j) \\
&= - \hat x_i \cdot ( \vec{v} \times \vec{\partial} \times \vec{p}).
\end{array}
\end{equation}
The vector notation version can be verified by writing out the terms, e.g.:
\begin{equation}
dp_1/dt - \partial_1 H(p) =  (v_2 (d_2 p_1 - d_1 p_2) + v_3 (d_3 p_1 - d_1 p_3))
\end{equation}
We have used an unconventional but more consistent notation $\vec{\partial}$. Inserting the more conventional notation $\nabla$, we have:
\begin{equation}
d\vec{p}/dt = \nabla H(p) - \vec{v} \times \nabla \times \vec{p}. 
\label{Hamiltonssecondlaw1}
\end{equation}
Which we consider to be the vector version of Hamilton's second equation of motion.

We can compare this with the Lorentz force law\cite{LandauLifshitz}---the electromagnetic force on a particle written in terms of the electric and magnetic fields or the scalar and vector potentials
\begin{equation}
\begin{array}{ll}
d\vec{p} / dt & =  e \vec{E} + e \vec{v} \times \vec{B}  \\
&=  e (- \nabla \Phi  - \partial_t \vec{A} ) + e \vec{v} \times (\nabla \times \vec{A})
\end{array}
\end{equation}
where we use the traditional notation for the electric and magnetic fields, $\vec{E}$, $\vec{B}$, and the scalar and vector potentials, $\Phi$, $\vec{A}$, respectively. 
The existence of a partial time derivative  $\partial_t A $ in electrodynamics requires comment. Thus far in ZM theory we have not used a partial time derivative since the time dependence arises from the space and clock field dependence. 
At this point, it seems natural to allow an explicit time dependence of the momentum and add a term $\partial_t \vec{p}$ to $d \vec{p}/dt$. While this appears possible, introducing an explicit time dependence, i.e. one different from the time dependence we have obtained from the space dependence, requires justification if it is to be used in ZM theory. Thus, in the meantime we do not make this assumption. Instead, we continue to assume that $\vec{p}$ has no explicit time dependence, and the explicit time dependence of $\vec{A}$ arises in electrodynamics from the choice of gauge in a manner that will become apparent shortly. 

We can identify directly a correspondence of Eq. (\ref{Hamiltonssecondlaw1}) with electromagnetism. A first correspondence would be
\begin{equation}
\begin{array}{ll}
e \Phi &\rightarrow - H(p(x))+H_0 \\
e \vec{A} &\rightarrow - \vec{p}(x) + \vec{p}_0
\end{array}
\label{emcorrespondence0}
\end{equation}
with $H_0$ and $\vec{p}_0$ constants. However, there is additional flexibility from the choice of gauge in electrodynamics. We define
\begin{equation}
\begin{array}{ll}
e \Phi &=  - H(p(x)) - \partial_t \xi(x,t) \\
e \vec{A} &= - \vec{p}(x) + \nabla \xi(x,t)
\end{array}
\label{emcorrespondence}
\end{equation}
where $\xi(x,t)$ an arbitrary function of space time, which can be included because it makes no contribution in the calculation of $d \vec p/dt$. Since we insert the function $\xi(x,t)$ in the correspondence with electromagnetism we do not assume that its arguments satisfy the properties of ZM theory space time. In the current derivation, this is the source of the explicit time dependence of the vector potential. The first ZM theory correspondence in Eq. (\ref{emcorrespondence0}) is thus a specific gauge choice and the more general correspondence in Eq.  (\ref{emcorrespondence}) enables the flexibility of choosing a gauge. Substituting, we show the correspondence of the electromagnetic equation to the ZM equations:
\begin{equation}
\begin{array}{ll}
d\vec{p} / dt & =  e (- \nabla \Phi  - \partial_t \vec{A} ) + e \vec{v} \times (\nabla \times \vec{A})  \\
& =   (- \nabla (- H(p(x)) - \partial_t \xi(x,t))  - \partial_t (- \vec{p}(x) + \nabla \xi(x,t)) )  +  \vec{v} \times (\nabla \times (- \vec{p}(x) + \nabla \xi(x,t)) )  \\
& =  \nabla H(p(x)) +  \nabla \partial_t \xi(x,t)   - \partial_t ( \nabla \xi(x,t))  - \vec{v} \times (\nabla \times \vec{p}(x)  ) \\
& =  \nabla H(p(x)) - \vec{v} \times (\nabla \times  \vec{p}(x) ).
\end{array}
\end{equation}
The first line is the Lorentz force equation given the scalar and vector potentials $\Phi$ and $\vec{A}$. The second line substitutes the correspondence we identified to ZM theory in Eq. (\ref{emcorrespondence}). From the second to the third line we collect terms and use the assumption that $\vec{p}(x)$ does not have an explicit time dependence, only a time dependence through spatial variation.  After cancellation, the final line is the ZM equation, Eq. (\ref{Hamiltonssecondlaw1}). 

\section{Total energy and canonical momentum}

For the one spatial dimension Hamilton's equation\cite{ZM2} we defined the total energy as the sum of the kinetic and potential energies in order to obtain a conserved total energy. This follows from the assumption that the changes in momentum experienced by a particle are due to external interactions that conserve the energy. For the case of three dimensions, we can similarly define a total energy that is conserved if the external forces are not explicitly time dependent:
\begin{equation}
H(p,x) = H(p) + e \Phi(x) 
\end{equation}
We also have a total momentum which corresponds to the canonical momentum.
\begin{equation}
\vec{\pi}(p,x) = \vec{p} + e \vec{A}(x)
\end{equation}
For the choice of correspondence given by Eq. \ref{emcorrespondence0}, without a time dependent term in the gauge, these are conserved quantities, and are equal to $H_0$ and $\vec{p}_0$ in Eq. \ref{emcorrespondence0}. Otherwise, they may be time dependent if there is a time dependent vector or scalar potential as a result of the gauge choice. Either way, the Lorentz force law continues to be valid. 

It is conventional to rewrite the Hamiltonian in terms of the canonical momentum as:
\begin{equation}
\begin{array}{ll} 
H(\pi,x) &= H( \pi-eA) + e \Phi(x)  \\
&= \sqrt{m^2+\sum_i \left(\pi_i-eA_i\right)^2}+ e \Phi(x).
\end{array}
\end{equation}
This completes the correspondence of our treatment with the conventional Hamiltonian treatment of the electromagnetic forces. We note, however, that it was not necessary to postulate the electromagnetic potentials and the form of the Hamiltonian. Instead, we obtained the form of the Lorentz force law, and the conserved energy and momentum from the non-orthogonality of space and time. We then substituted into the Hamiltonian to obtain its conventional form including the electromagnetic potentials assuming conservation laws hold.

In the conventional treatment of the Hamiltonian $p_i$ is a fundamental coordinate which has no explicit space dependence. In deriving the equations of motion, the space and time dependence of the potentials $A_i$ and $\Phi$ provide the spatial dependence. Explicitly, in the traditional formalism $\partial_j p_i = 0 $ and without the potentials $\partial_i H = 0$. The incorporation of the potentials in the Hamiltonian results in the correct values of the derivatives to obtain the equations of motion from Hamilton's second law. In the ZM theory derivation, the spatial dependence is already present in $p_i(x)$, reflecting the spatial properties of the clock field. In order to make the correspondence to the traditional picture, we introduce the potentials assuming that the canonical momentum and total energy are independent of position.  The time dependence can be considered in the same way except that the question of an extrinsic time dependence requires further discussion.

\section{$E$ and $B$ as space-time coordinate transformations}

For further development of ZM theory it may be helpful to interpret the electromagnetic fields in terms of coordinate transformations. For this we consider the spatial variation of $\vec{p}(x)$ as seen by an observer to be due to coordinate transformations of space relative to the observer. The value of the momentum reflects a difference between an observer frame of reference, and the co-moving frame of reference at a location, $x$. Thus, we can adopt the point of view that without coordinate transformations, $\vec{p}(x)$ would be uniformly zero. A uniform value of $\vec{p}(x)$ results from a single coordinate transformation of all of space, while a spatially varying value of $\vec{p}(x)$ results from spatial variation of coordinate transformations.   

Considering the spatial variation of  $\vec{p}(x)$ and its relation to $\vec{B}$ and $\vec{E}$ we see that $\vec{B}$ is the local rotation of space, and $\vec{E}$ the bending of the direction of time (analogous to boosts in special relativity). 

Explicitly, $e\vec{B}$ is equal to $-\nabla \times \vec{p}$ (there is no gauge contribution to this expression). This means that $B$ is the local rotation of the $\vec{p}(x)$ field. Alternatively, assuming that there is no inherent rotation of $\vec{p}(x)$, only a perceived rotation due to relative transformation of coordinates to the observer, this is the rotation of space relative to the observer at that space location. 

$\vec{E}$
is equal to $\nabla H(p(x))$, which means that $\vec{E}$ is related to the bending of time ---  it is the local change of the secant of the direction of time. 

\section{comments}

We note that the derivation of the Lorentz force law demonstrates a consistency of formalism between ZM theory and electrodynamics for the action of fields on matter. Demonstrating correspondence to electrodynamics also requires obtaining the source equations of the fields. In this paper we have assumed the spatial variation of the momentum and energy. It remains to be shown that this variation is consistent with the source equations of electrodynamics. 

ZM theory appears to fix the choice of gauge, as it fixes the choice of coordinate system. It is possible to view this as due to the assumption that the local velocity is given by the clock field momentum, e.g. in Eq. \ref{first} and Eq. \ref{second}. If we allow the existence of multiple clock fields, the gauge obtained may not be the same for each clock field. 

We have found that in ZM theory the momentum at a point in space is linked to the electromagnetic field at that point. One of the conceptual difficulties with ZM theory in relation to traditional classical physics is the conventional treatment of the motion of particles in fixed fields that comprises a significant component of textbooks.\cite{Jackson} This approach leads to the impression that a fixed field determines particle acceleration at a point in space (a ``test" particle), and allows its momentum or velocity to be arbitrary at that point. However, while these fixed field treatments are useful, fields are not independent of particle momentum or velocity in classical electrodynamics. An example would be a classical treatment of a central potential with orbits that are considered similar to those in a Newtonian gravitational central potential. This picture is not a complete one in classical electrodynamics.\cite{Jackson} If we consider two oppositely charged particles orbiting around their center of mass, both particles contribute to the electromagnetic fields. If one particle is more massive than the other, the motion of the massive particle, which depends on the motion of the less massive particle, also affects the fields it generates. Further, the orbital acceleration results in radiative dissipation of energy, and the orbit is not stationary. The proper way to address self-interaction in this context presents essential difficulties that are not overcome in classical electrodynamics and are linked to quantum field theoretic issues. This leaves open the possibility of considering the relationships between fields and momentum embedded in ZM theory. 

We note that gauge theory generalizations of quantum electrodynamics, are based upon expanding the gauge symmetry from $U(1)$ to larger and non-Abelian groups. The utility of generalizing the clock field from $U(1)$ remains to be discussed.

\section{summary}

In this paper we extended the analysis of correspondence of ZM theory to traditional theory by deriving the Lorentz force law directly from the assumption of the non-orthogonality of space and time in ZM theory. The derivation did not make assumptions about the nature of electric and magnetic fields, it considered directly the ZM clock field behavior in three spatial dimensions. The correspondence enabled identification of the electric and magnetic fields, and the electromagnetic scalar and vector potentials, in terms of the spatial variation (momentum) of the clock field---the clock field on which they act. This identification also associated the electric and magnetic fields with space-time coordinate transformations.

I thank Marcus A. M. de Aguiar for helpful comments on the manuscript.

\end{document}